\documentclass[sigconf]{acmart}
\AtBeginDocument{%
  }

\copyrightyear{2025}
\acmYear{2025}
\setcopyright{acmlicensed}\acmConference[MM '25]{Proceedings of the 33rd ACM International Conference on Multimedia}{October 27--31, 2025}{Dublin, Ireland}
\acmBooktitle{Proceedings of the 33rd ACM International Conference on Multimedia (MM '25), October 27--31, 2025, Dublin, Ireland}
\acmDOI{10.1145/3746027.3767108}
\acmISBN{979-8-4007-2035-2/2025/10}

\begin{document}

\title{Fact-Checking at Scale: Multimodal AI for Authenticity and Context Verification in Online Media}
\author{Van-Hoang Phan}
\authornote{These authors contributed equally to this research.}
\orcid{0009-0003-8615-435X}
\affiliation{
  \institution{University of Science}
  \city{Ho Chi Minh City}
  \country{Vietnam}
}
\email{21120459@student.hcmus.edu.vn}

\author{Tung-Duong Le-Duc}
\authornotemark[1]
\orcid{0009-0001-2603-0908}
\affiliation{
  \institution{University of Science}
  \city{Ho Chi Minh City}
  \country{Vietnam}
}
\email{23125081@student.hcmus.edu.vn}

\author{Long-Khanh Pham}
\authornotemark[1]
\orcid{0009-0004-6197-3428}
\affiliation{
  \institution{FPT Software AI Center}
  \city{Ho Chi Minh City}
  \country{Vietnam}
  \institution{University of Science}
  \city{Ho Chi Minh City}
  \country{Vietnam}
}
\email{21120479@student.hcmus.edu.vn}

\author{Anh-Thu Le}
\orcid{0009-0001-4160-4746}
\affiliation{
  \institution{VinUni-Illinois Smart Health Center, VinUniversity}
  \city{Hanoi}
  \country{Vietnam}
}
\email{thu.la2@vinuni.edu.vn}

\author{Quynh-Huong Dinh-Nguyen}
\orcid{0009-0004-1162-7770}
\affiliation{
  \institution{University of Science}
  \city{Ho Chi Minh City}
  \country{Vietnam}
}
\email{22127146@student.hcmus.edu.vn}

\author{Dang-Quan Vo}
\orcid{0009-0005-8031-082X}
\affiliation{
  \institution{University of Science}
  \city{Ho Chi Minh City}
  \country{Vietnam}
}
\email{23125090@student.hcmus.edu.vn}

\author{Hoang-Quoc Nguyen-Son}
\orcid{0000-0003-2468-7815}
\affiliation{
  \institution{National Institute of Information \\ and Communications Technology}
  \city{Tokyo}
  \country{Japan}
}
\email{quoc-nguyen@nict.go.jp}

\author{Anh-Duy Tran}
\orcid{0000-0002-8036-954X}
\affiliation{
  \institution{DistriNet, KU Leuven}
  \city{Leuven}
  \country{Belgium}
}
\email{anh-duy.tran@kuleuven.be}

\author{Dang Vu}
\authornote{Corresponding authors.}
\orcid{0009-0003-3268-024X}
\affiliation{
  \institution{FPT Software AI Center}
  \city{Ho Chi Minh City}
  \country{Vietnam}
  \institution{University of Science}
  \city{Ho Chi Minh City}
  \country{Vietnam}
}
\email{dangvqm@fpt.com}

\author{Minh-Son Dao}
\authornotemark[2]
\orcid{0000-0003-3044-8175}
\affiliation{%
  \institution{National Institute of Information \\ and Communications Technology}
  \city{Tokyo}
  \country{Japan}
}
\email{dao@nict.go.jp}

\renewcommand{\shortauthors}{Van-Hoang Phan et al.}

\begin{abstract}
The proliferation of multimedia content on social media has transformed how information is produced and consumed, enabling real-time coverage of global events but also accelerating the spread of misinformation, particularly during crises such as wars, natural disasters, and elections. The rise of synthetic media and the reuse of authentic content in misleading contexts further underscore the need for robust verification tools. In this paper, we present a comprehensive system developed for the ACM Multimedia 2025 Grand Challenge on Multimedia Verification. Our system evaluates both the authenticity and contextual accuracy of multimedia content in multilingual settings, producing expert-oriented verification reports alongside accessible summaries for the public. We propose a unified verification pipeline that integrates visual forensics, textual analysis, and multimodal reasoning, with a hybrid approach to detecting out-of-context (OOC) media via semantic similarity, temporal alignment, and geolocation cues. Extensive evaluations on the challenge benchmark demonstrate the system’s effectiveness across diverse real-world scenarios. Our contributions advance the state of the art in multimedia verification while providing practical tools for journalists, fact-checkers, and researchers addressing information integrity in the digital age.
\end{abstract}



\maketitle

\section{Introduction}
\label{sec: Introduction}

Social media has become a primary channel for news and information, but the rapid circulation of multimedia content increases risks of misinformation and manipulation. Multimedia verification, now central to digital fact-checking, evaluates the authenticity, origin, and integrity of visual content. Advances in Open Source Intelligence (OSINT) tools have strengthened verification, yet challenges remain in handling multimodal data, multilingual content, and detecting subtle context shifts or AI-generated artifacts.

In this paper, we present our system developed for the ACM Multimedia 2025 Grand Challenge on Multimedia Verification \cite{MV2025overview}. The challenge addresses the growing problem of misinformation and disinformation in online multimedia, where authentic and synthetic content are often intermingled and reused out of context. Our system is designed to assess the authenticity and contextual accuracy of multimedia content in multilingual settings, with a special focus on the Out-of-Context (OOC) sub-task. It produces structured, interpretable reports for fact-checkers as well as simplified summaries for general users.

Our main contributions are:
\begin{itemize}
    \item A unified multimedia verification pipeline integrating visual forensics, text analysis, and multilingual support, including a hybrid OOC detection method that combines semantic, temporal, and geolocation cues.
    \item An interpretable reporting design tailored for both expert analysts and non-expert users, evaluated through the Grand Challenge framework.
\end{itemize}

\section{Proposed Method}
\label{sec: Proposed_Method}
The multimedia verification framework employs a multi-stage pipeline designed to comprehensively analyze multimedia content through evidence aggregation and contextual analysis. The system integrates four core verification services that address the fundamental questions of multimedia authenticity: geolocation ("Where?"), temporal analysis ("When?"), content classification, and intent analysis ("Why?"). The framework processes diverse input modalities, including JSON metadata, images, and videos, converting them into standardized formats for downstream analysis services.

The overall architecture follows a unified approach that combines visual forensics, natural language processing, and web-based evidence gathering to produce structured verification reports. The system leverages large language models for content classification and reasoning, while employing specialized deep learning models for geolocation and temporal analysis. Each verification service operates independently but contributes to a comprehensive evidence aggregation module that synthesizes findings into interpretable reports for fact-checkers and end users.

\begin{figure}[t]
    \centering
    \includegraphics[width=0.4\textwidth]{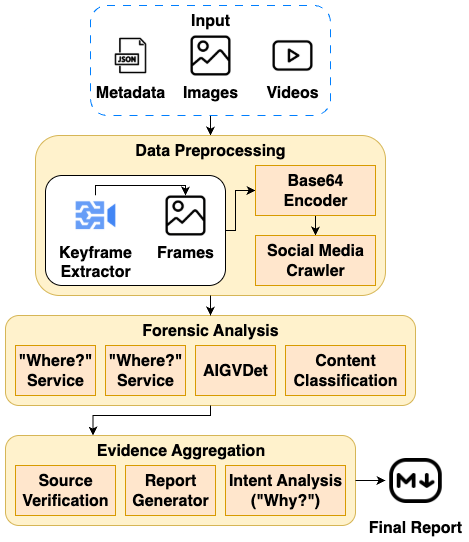}
    \caption{The pipeline begins with three types of inputs: JSON metadata, images, and videos. In the data preprocessing stage, metadata is processed via a social media crawler, transcripts are extracted from video content using OpenAI Whisper, while images and video frames—extracted using Google Vision Intelligent keyframe extraction—are encoded in Base64 format for standardized handling. The preprocessed data is then routed to four specialized verification services: "Where?" (geolocation), "When?" (temporal analysis), content classification, and AIGVDet (AI-generated visual detection). Outputs from these services are aggregated in the evidence aggregation module, which performs source verification, intent analysis ("Why?"), and report generation. The final output is a structured verification report designed to support fact-checking and multimedia forensics.}
    \label{fig:overall}

\end{figure}

\subsection{Data Preprocessing}

The data preprocessing pipeline handles heterogeneous multimedia inputs through a systematic three-stage process: metadata extraction, media processing, and format standardization.

\begin{itemize}
    \item \textbf{Metadata Processing:} The system processes JSON metadata files containing structured information including title, description, location, category, violence level, and social media links. This metadata serves as contextual information for subsequent verification services and provides initial cues for content classification and temporal analysis.
    
    \item \textbf{Media Encoding:} All processed media content undergoes Base64 encoding to enable standardized API transmission across verification services. The system implements separate encoding functions for static images and video frames, ensuring consistent format handling regardless of input type.

    \item \textbf{Integration and Standardization:} The preprocessing module integrates metadata and processed media into a unified data structure compatible with downstream verification services. This includes organizing media content, extracting relevant textual features, and preparing data payloads for geolocation analysis, temporal verification, and content classification services. The standardized output format enables seamless integration with the evidence aggregation pipeline while maintaining data integrity throughout the verification.
\end{itemize}

\subsection{Content Classification}
\label{subsec:category}
Content classification is a crucial step in organizing and understanding the verified multimedia content. This process aims to categorize the content by assigning relevant tags based on its subject matter, origin, and key entities involved. Large Language Models (LLMs) are employed for this task, leveraging their advanced semantic understanding to generate a comprehensive set of keywords. These tags are derived from the aggregated textual evidence gathered from supporting sources. The generated tags span various categories, including the social media platforms where the content circulated (e.g., TikTok, Telegram, X), prominent individuals or organizations depicted or mentioned (e.g., Trump, Coca-Cola, Hamas, Al Jazeera), and specific events or topics to which the content pertains (e.g., Ukraine War, AI-generated, Gaza, Shifa Hospital). This systematic tagging facilitates efficient retrieval and contextualization of verified information.

\subsection{Verified Evidence}
\label{subsec:evidence}

\subsubsection{Source Details}
\label{subsubsec:details}
By analyzing the gathered evidence using LLM, the system meticulously identifies key attributes for each supporting source. This includes pinpointing the original publisher or entity responsible for the content, specifying the platform on which it appeared (e.g., social media, news website), and providing direct links (URLs) to the original posts or articles. Additionally, any available publication dates and timestamps are extracted. The aim is to present a clear, traceable lineage of the content, enabling fact-checkers to assess its primary distribution and initial context effectively.

\subsubsection{Where? (Location)}
\label{subsubsec:where}
\begin{figure*}[h]
    \centering
    \includegraphics[width=.95\textwidth]{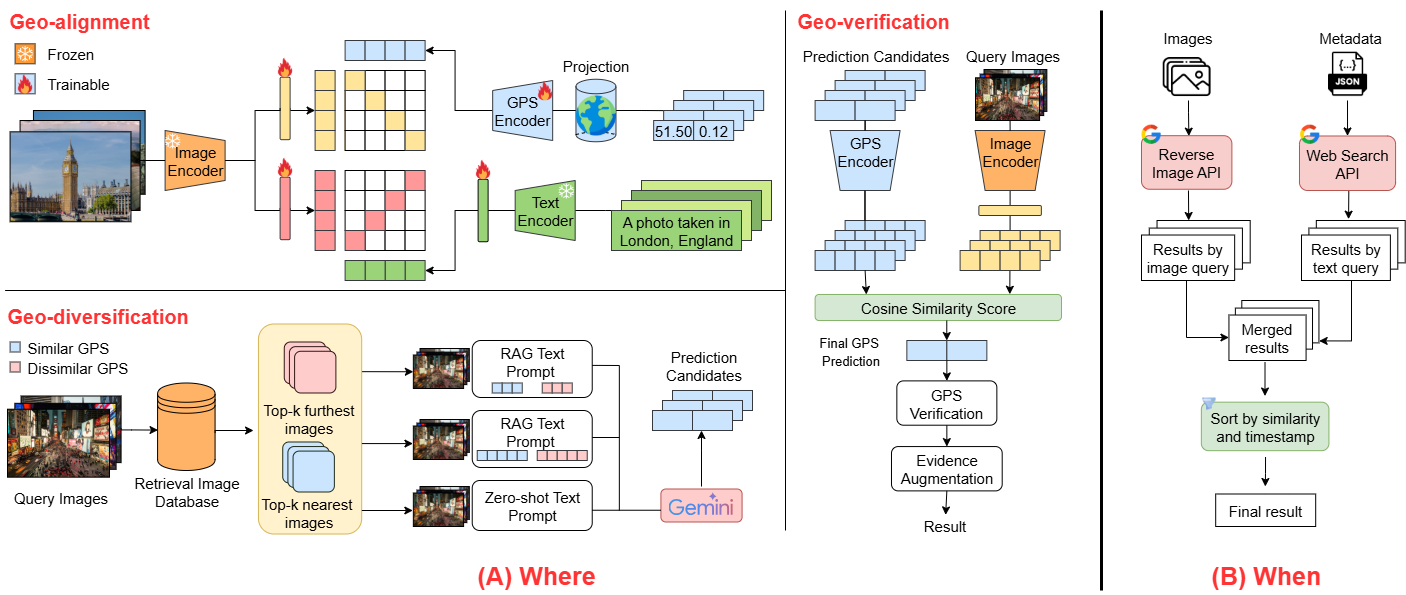}
    \caption{The pipeline of finding (A) where and (B) when images/ videos are taken}
    \label{fig:where-when-pipeline}
\end{figure*}

To estimate the geographic locations of images and videos, an enhanced version of the G3 framework~\cite{jia2024g3} is adopted, as illustrated in Figure~\ref{fig:where-when-pipeline}. The pipeline comprises two main phases: Database Construction and Location Prediction.

\textbf{Phase 1: Database Construction.} This phase builds an image database using the pretrained model obtained from the Geo-alignment process, in order to support the large multimodal model (LMM) in the next phase.

\begin{itemize}
    \item \textbf{Geo-alignment.} In this step, images, textual descriptions, and GPS coordinates are aligned within a shared embedding space. Visual and textual features are extracted using frozen CLIP encoders, while GPS coordinates are processed through a Mercator projection followed by Random Fourier Feature encoding. All representations are jointly trained using contrastive loss.

    \item \textbf{Image Database.} The database is constructed from the MP16-Pro dataset~\cite{jia2024g3}. Each image is embedded using the pretrained image encoder, and the resulting visual features are combined with corresponding textual and GPS-aligned representations to retain geolocation information.
\end{itemize}

\textbf{Phase 2: Location Prediction}. To estimate the final GPS coordinates and location, two key steps from G3~\cite{jia2024g3} are used: Geo-diversification and Geo-verification.
\begin{itemize}
    \item \textbf{Geo-diversification.} The system leverages Google's Gemini Pro 2.5 as the LMM for a Retrieval-Augmented Generation (RAG) pipeline. To enable LMM to reason over rich contextual information, the input prompt includes similar and dissimilar GPS coordinates, metadata, available transcripts, and search results retrieved from Cloud Vision Web Detection and Scrapingdog. The LMM processes multiple such prompts to generate candidate predictions, each consisting of a GPS estimate, the corresponding location, and supporting evidence.
    
    \item \textbf{Geo-verification.} This process starts by encoding each predicted GPS and computing their cosine similarity against the input images. The prediction with the highest average similarity score is then selected, and its location is verified by the LMM and OpenStreetMap's Nominatim service. Finally, the supporting evidence is augmented with images of the predicted location, ensuring that each piece of evidence is supported by input images, relevant location images, and URLs.
\end{itemize}

\subsubsection{When? (Time)}
\label{subsubsec:when}


To detect the capture or recording time of multimedia content (images/videos), we propose a practical pipeline that leverages metadata, vision-language modeling, and web search engines.

Given an input multimedia file $M$ (image or video) and an associated metadata file $J$ (in JSON format), the system outputs:
\begin{itemize}
    \item \textbf{Timestamp:} The estimated capture or recording time of $M$.
    \item \textbf{Source:} A relevant webpage or article containing $M$ or similar visual evidence.
    \item \textbf{Confidence:} The similarity score between metadata and textual content of the source.
\end{itemize}

Two types of search are performed:
\begin{itemize}
    \item \textbf{Textual Search:} A search query is constructed from the metadata:
    \[
    q = \texttt{location} + \texttt{title} + \texttt{description}
    \]
    and submitted to \textbf{Google Web Search API} via SerpAPI.

    \item \textbf{Reverse Image Search:} For each keyframe (or the image input), the image is uploaded to imgbb to obtain a public URL, which is then used in \textbf{Google Reverse Image Search} via SerpAPI.
\end{itemize}

The results of both search types are merged.

Each result is scored using:
\begin{itemize}
    \item \textbf{Textual similarity:} Using SequenceMatcher between the result’s title+snippet and the metadata query $q$.
    \item \textbf{Timestamp parsing:} Dates are extracted and normalized to standard formats.
\end{itemize}

Results are sorted in descending order of similarity. Ties are broken using ascending publication date. Due to cost constraints, the system adopts an \textit{early stopping strategy}: the first high-confidence result containing a valid timestamp terminates the search process.

To comply with the limited query quota of SerpAPI (100 searches
/month on the free tier), we employ an early stopping mechanism that halts the search once a confident timestamp is found. Although this reduces cost, it may overlook stronger evidence in later keyframes. A fully exhaustive approach—processing all keyframes and ranking all results—would improve reliability but requires more query budget. Future improvements may include adaptive stopping or cost-aware scoring policies.


\subsubsection{Why? (Motivation or Intent)}
\label{subsubsec:why}
In our methodology, we leverage the enhanced reasoning and explainability capabilities of large language models (LLMs), facilitated by advanced prompting strategies such as chain-of-thought. Specifically, our system constructs a reasoned explanation of the model’s intent by aggregating intermediate evidence generated from earlier processing stages and applying multi-level reasoning. This structured reasoning process supports the derivation and justification of the final decision output.

\paragraph{Evidences Gathering}
The gathering of evidence systematically acquires relevant external information to augment the multimedia verification process. The approach begins with a reverse image search that utilizes the input media (or extracted keyframes from video) to identify corresponding and partially matching web pages. The URLs thus identified are subsequently processed to extract contextual information. For standard Web articles, core content, titles, authors, and publication dates are extracted via an automated parsing mechanism. Specialized handling is used for social media platforms, such as X (formerly Twitter), where post content and metadata are retrieved through a dedicated scraping approach. All external links collected and their corresponding textual content are aggregated, establishing a comprehensive pool of supporting evidence for subsequent analysis stages, including content classification and intent analysis.

\subsection{Forensic Analysis}
\label{subsec:forensic}

To assess video authenticity, we adopt the AIGVDet model~\cite{AIGVDet}, a forensic framework designed to detect AI-generated content. AIGVDet comprises two key modules: a Spatial Domain Detector and an Optical Flow Detector.

The Spatial Domain Detector analyzes individual RGB frames to identify visual artifacts such as unnatural textures, lighting inconsistencies, and anomalous pixel patterns. Features are extracted using a ResNet50 backbone~\cite{resnet50}. In parallel, the Optical Flow Detector captures temporal artifacts using flow maps computed via the RAFT algorithm~\cite{raft}, exploiting the fact that generative models often fail to preserve coherent motion.

Let \( I_i \) denote the \( i \)-th frame, and \( F_i = \mathcal{F}(I_i, I_{i+1}) \) the corresponding optical flow. Features \( V_i^I \) and \( V_i^F \) are extracted using ResNet50 encoders \( R(\cdot) \) and \( O(\cdot) \):

\begin{equation}
    V_i^I = R(I_i), \quad V_i^F = O(F_i)
\end{equation}

These features are passed through a global average pooling layer and a fully connected layer with sigmoid activation:

\begin{equation}
    P_i^I = \sigma(\mathrm{FC}(\mathrm{GAP}(V_i^I))), \quad 
    P_i^F = \sigma(\mathrm{FC}(\mathrm{GAP}(V_i^F)))
\end{equation}

The fused frame-level prediction is:

\begin{equation}
    P_i = \alpha P_i^I + (1 - \alpha) P_i^F
\end{equation}

The final video-level score is the average across all frames:

\begin{equation}
    P = \frac{1}{N - 1} \sum_{i = 1}^{N - 1} P_i
\end{equation}

AIGVDet is trained on the Generated Video Dataset (GVD), which includes synthetic videos from 11 generative models spanning both text-to-video and image-to-video pipelines~\cite{AIGVDet}. In our pipeline, we adopt AIGVDet not for binary classification, but to produce confidence scores for both \textit{authentic} and \textit{synthetic} labels. This probabilistic output provides a more nuanced and interpretable assessment of video authenticity.

\subsection{Summary of Key Points}
\label{subsec:summary}
Our system leverages Large Language Models (LLMs) to synthesize a concise and comprehensive case summary from all available verification evidence. This includes initial textual supporting sources, contextual media information, and detailed findings from the "When?", "Where?", "Who?", and "Why?" analysis modules. The LLM is guided by a structured prompt and a defined JSON schema to ensure the output is consistent, accurate, and readily interpretable. The generated summary integrates findings across all modalities, highlighting the central claim or event, primary verification outcomes, and any identified uncertainties or unknowns.

\subsection{Sub-Task: OOC Detection}
\label{subsec:ooc-detection}

\subsubsection{SearchOOC}

\begin{figure}[ht]
    \centering
    \includegraphics[width=.9\linewidth]{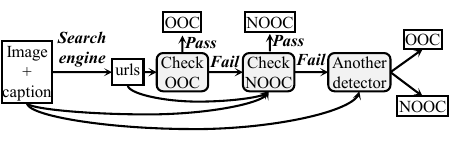}
    \caption{SearchOOC for detecting out-of-context}
    \label{fig:SearchOOC}
\end{figure}

We propose a method called SearchOOC for detecting whether an image-caption pair is out-of-context (OOC) or not (Figure~\ref{fig:SearchOOC}):
\begin{enumerate}
    \item \textbf{Search for Evidence:} We use Google Search with the input image-caption pair to obtain a list of URLs that likely contain both the image and the caption.
\item \textbf{Check for OOC:} Among the retrieved URLs, we check whether any site explicitly mentions that the pair is out-of-context. If we find a website making this determination, SearchOOC classifies the input pair as OOC. Otherwise, we proceed to the next step.
\item \textbf{Check for NOOC:} We check whether any of the URLs belong to a well-known website that restricts access (such as The New York Times). If so, we classify the input pair as not out-of-context (NOOC). If not, we further verify if any website actually contains the input image and caption. Since the image might be resized and the caption paraphrased, we use image hashing (with a threshold of 10) and text similarity (with a threshold of 7.5) to make this determination. If either check passes, we classify the pair as NOOC. If not, we move to the next step.
\item \textbf{Detect Using an Alternative Method:} In this step, we apply a baseline model (o4-mini) with the following direct prompt: \textit{``Given an image and the caption, determine whether the image and caption are out of context. Please output only `ooc' or `nooc' without any explanation.''}
 
\end{enumerate}
 
\subsubsection{HierOOC}

HierOOC is a hierarchical, multi-stage framework designed to detect whether a given input—including an image, a caption, and optionally a context—is out-of-context (OOC). The system consists of four iterative stages:

\begin{enumerate}
    \item \textbf{Multi-scale Evidence Retrieval:} Given the input pair, this stage retrieves evidence at two levels. At the \textit{external scale}, a search engine is used to collect relevant URLs and news articles. At the \textit{internal scale}, an entity detector extracts textual and visual entities from the image-caption pair.

    \item \textbf{Evidence Filtering:} To ensure evidence quality and relevance, HierOOC applies several filtering strategies, including domain filtering, language filtering, redundancy removal, and similarity-based filtering. A similarity threshold of 0.7 is used to retain only the most relevant candidates.

    \item \textbf{Evidence Refinement:} Some retrieved evidence, particularly from web sources, may still be inconsistent with the original input and internal candidates. Therefore, a large language model (LLM) is employed to assess and eliminate inconsistencies, thereby reducing the input token size for subsequent stages instead of passing all collected evidence forward.

    \item \textbf{OOC Decision Making:} In the final stage, the OOC Decisor determines whether the input is out-of-context. It takes the input pair and the refined evidence summary as input. This module includes detection agents guided by Chain-of-Thought (CoT) prompting to improve reasoning and reduce hallucination. All LLMs used in this framework are based on the Gemini 2.5 Flash Lite version.
\end{enumerate}

\begin{figure}[ht]
    \centering
    \includegraphics[width=\linewidth]{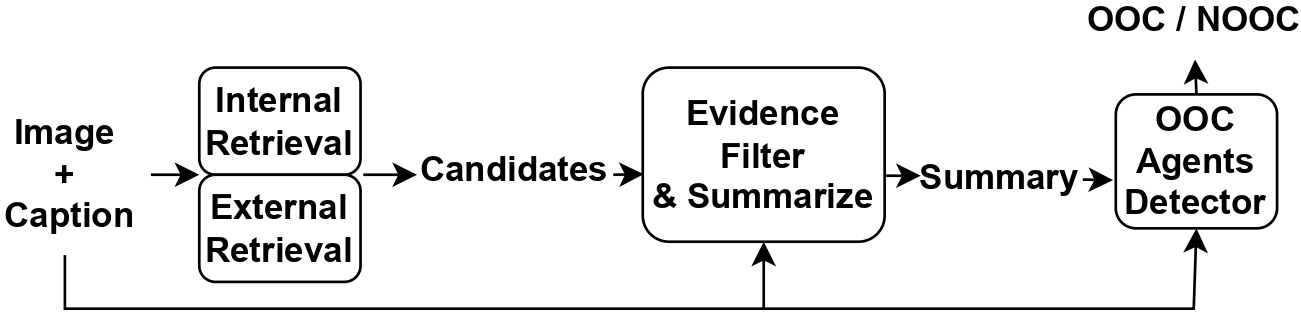}
    \caption{HierOOC for detecting out-of-context}
    \label{fig:HierOOC}
\end{figure}

\section{Experimental Result}
\label{sec: Experiments}

\subsection{Dataset}
\subsubsection{Main-task}
The main-task dataset contains real-world multimedia cases, each with images or videos, multilingual context (such as captions, posts, or articles), and metadata. Drawn from fact-checking archives, the dataset reflects authentic events and may include sensitive material. For each case, participants must analyze the content and submit an English verification report summarizing findings, content categories, forensic analysis, and key evidence including source, location, time, and involved entities.
\subsubsection{Sub-task}
\label{subsec: dataset}
We use the COSMOS dataset \cite{aneja2021cosmos} for our experiments. COSMOS is a recently compiled benchmark for Cheapfakes detection. It consists of images and captions scraped from news articles and other websites, designed to capture the manipulation or misuse of multimedia content by altering the context of an image or video with misleading captions or claims. COSMOS contains 161,752 images for training, 41,006 images for validation, and 1,000 images for testing. Each image is associated with multiple captions, some of which are in-context and some are out-of-context. The test set is manually annotated with binary labels indicating whether a given image-caption pair is in-context or out-of-context. 

\subsection{Sub-task Metrics}
\label{subsec: metrics}
To evaluate the effectiveness of the proposed method, the challenge use accuracy as the main criterion. 
The formula for calculating accuracy is as follows:
$$accuracy = \dfrac{TP + TN}{TP + TN + FP + FN}$$
where $TP$ is the number of samples classified as OOC (out-of-context), $TN$ is the number of samples classified as NOOC (not-out-of-context), $FP$ is the number of samples misclassified as OOC and $FN$ is the number of samples misclassified as NOOC.

\subsection{Results}
\label{subsec: results}
 In this section, we present the results of our experiments on the cheapfake detection task using accuracy as the metric. We compare our proposed model with several methods on the public test of the COSMOS dataset.

\subsubsection{Main-task}

\begin{table}[htbp]
\centering
\caption{Leaderboard (Total Accuracy Scores) for the Grand Challenge}
\begin{tabular}{lcccc}
\hline
\textbf{} & \textbf{Team 1} & \textbf{Team 2} & \textbf{Team 3} & \textbf{Ours} \\
\hline
\textbf{Score} & \textbf{844.22} & 487.19 & 295.86 & \underline{644.65} \\
\hline
\end{tabular}
\label{tab:maintask}
\end{table}

As shown in Table \ref{tab:maintask}, our system (Ours) achieved a total accuracy of 644.65, securing a clear lead over other participating teams. The margin of more than 150 points compared to the next-highest competitor demonstrates the robustness and adaptability of our verification pipeline across diverse evaluation cases. These results highlight the effectiveness of our approach in maintaining high performance throughout the challenge and its strong competitiveness in the task setting.

\begin{figure}[ht]
    \centering
    \includegraphics[width=0.8\linewidth]{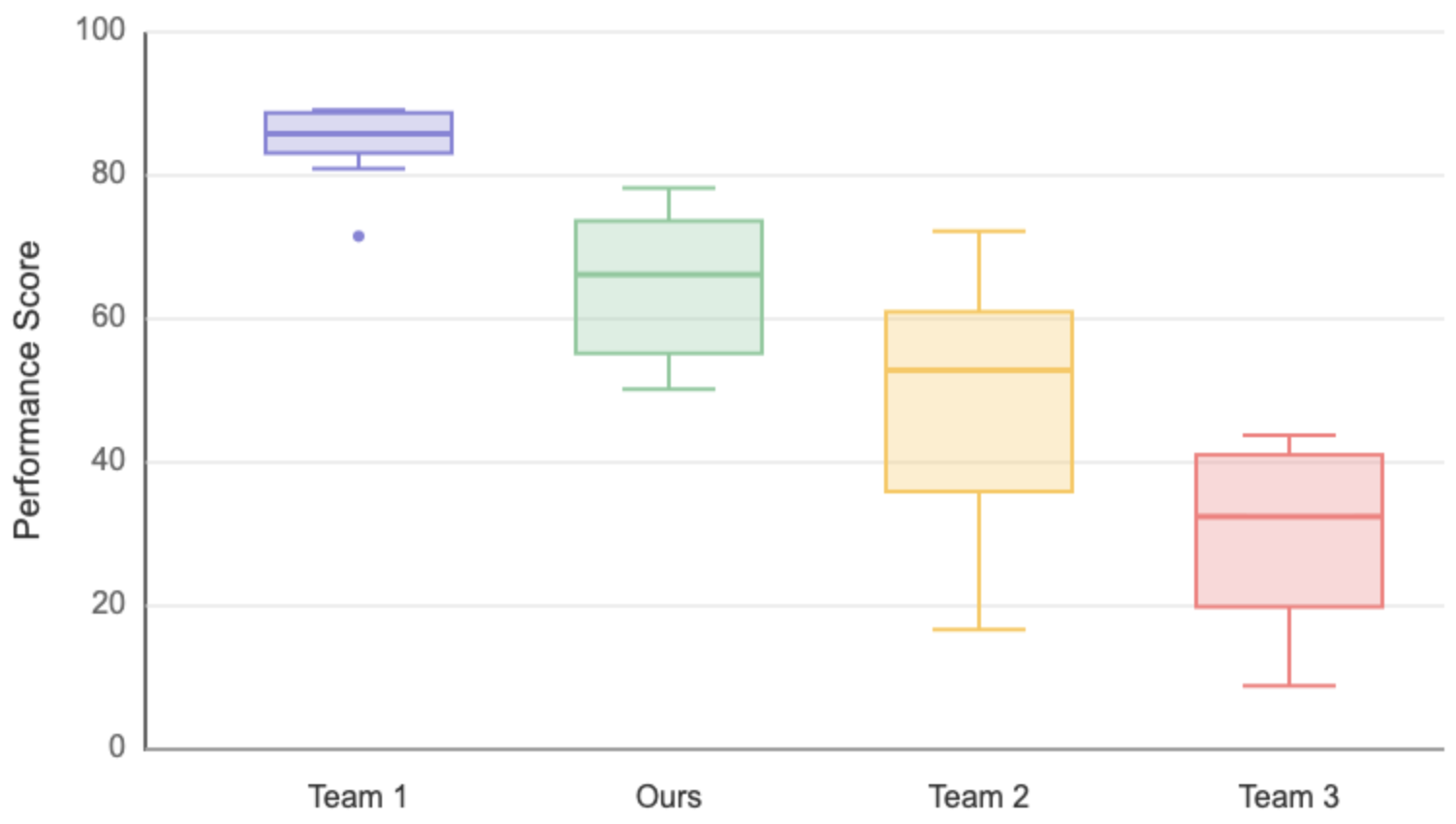}
    \caption{Performance and Stability Comparison of Four Methods Across Ten Test Cases}
    \label{fig:box-result}
\end{figure}

Figure ~\ref{fig:box-result} compares the performance and stability of four methods across ten test cases using box plots. Team 1 achieved the highest performance with exceptional stability (median = 85.0, IQR = 5.6), while our method demonstrated moderate performance with good consistency (median = 64.5, IQR = 18.5). Team 2 showed high variability (range = 55.6) and Team 3 exhibited the poorest performance with substantial instability (median = 29.6). The narrow interquartile ranges for Team 1 and our method indicate superior algorithmic stability compared to the wider distributions observed in Teams 2 and 3.

\subsubsection{Sub-task}


\begin{table}[h]
 \centering
 \caption{Sub-task's Performance Comparisons}
\begin{tabular}{ll||ll}
\toprule
\textbf{Method}      & \textbf{Acc(\%)} & \textbf{ Method}      & \textbf{Acc(\%)} \\
\midrule
Le et al.~\cite{le2024tega} & 56.4&Nguyen et al.~\cite{nguyen2024unified} &    93.0\\
Tran et al.~\cite{tran2022textual} & 73.0& o4-mini& 86.2\\
La et al.~\cite{la2022combination}& 76.0& \textbf{HierOOC}    & \textbf{95.3}   
\\
Dao et al.~\cite{dao2023leveraging} & 84.0& \textbf{SearchOOC}        & \textbf{96.0 }     \\
\end{tabular}
\label{tab:res}
\end{table}



HierOOC and SearchOOC demonstrate the effective integration of large language models (LLMs) with external information retrieved from search engines to enhance out-of-context (OOC) detection. Existing approaches, which rely solely on internal model knowledge, often lack sufficient contextual information to accurately determine whether an input pair is OOC or not. This limitation is consistent with findings by Luo et al.~\cite{luo2021newsclippings}, who conducted a study involving human evaluators restricted from using search engines, resulting in an average accuracy of only around 65\%. Notably, the o4-mini model achieves performance comparable to existing methods, highlighting the strength of modern LLMs. Furthermore, SearchOOC, which leverages external information with o4-mini as its backbone, improves performance in OOC detection.

\section{Conclusion}
\label{sec: Conclusion}

In this paper, we introduced a unified and comprehensive system for multimedia authenticity and context assessment, developed to address the pressing challenges of misinformation and disinformation on social media platforms. Our system tackles the complex task of verifying content across diverse modalities, including images, videos, and text, in multilingual environments. Key contributions include a multi-stage pipeline integrating visual forensics, textual analysis, and robust multimodal reasoning. We specifically highlighted our novel hybrid approaches, SearchOOC and HierOOC, for effectively detecting out-of-context media by combining semantic similarity, temporal alignment, and geolocation cues.

Extensive evaluations on the ACM Multimedia 2025 Grand Challenge benchmark \cite{MV2025overview}, particularly for the challenging out-of-context sub-task, demonstrate the strong effectiveness of our proposed methods. By generating both detailed expert-oriented verification reports and accessible public summaries, our system provides practical, interpretable tools vital for journalists, fact-checkers, and researchers. This work significantly advances the state of the art in multimedia verification, paving the way for more reliable information ecosystems and enhancing public trust in the digital age. Future work will focus on continuous improvement of cross-modal alignment, enhancing robustness against novel manipulation techniques, and expanding evaluation across a wider array of real-world datasets and languages.


\begin{acks}

All intellectual property rights arising from the work shall belong solely and exclusively to the National Institute of Information and Communications Technology (NICT), and may be used by NICT for any purpose, including but not limited to commercial applications, without any claim, restriction, or obligation from or to non-NICT contributors.
\end{acks}

\newpage

\bibliographystyle{ACM-Reference-Format}
\bibliography{ref}

\appendix

\end{document}